\documentclass[12pt,preprint]{aastex}
\usepackage{graphicx}
\usepackage{txfonts}
\usepackage{natbib}
\bibpunct{(}{)}{;}{a}{}{,}
\newcommand{\hd}{HD~189733}
\begin{document}

\title{XMM-Newton observations of \hd\ during planetary transits.}

\author{I. Pillitteri\altaffilmark{1} \and S. J. Wolk\altaffilmark{1} 
\and O. Cohen\altaffilmark{1}  \and V. Kashyap\altaffilmark{1} 
\and H. Knutson\altaffilmark{1,2} 
\and C. M. Lisse\altaffilmark{3}
\and G. W. Henry\altaffilmark{4}} 
\affil{SAO-Harvard Center for Astrophysics, 60 Garden St, Cambridge MA 02139 USA}
\affil{Department of Astronomy, University of California, Berkeley, 601 Campbell Hall, Berkeley CA 94720}
\affil{Planetary Exploration Group, Space Department, Johns Hopkins University Applied Physics Laboratory, 11100 Johns Hopkins Rd, Laurel, MD 20723  }
\affil{Tennessee State University, Center of Excellence in Information Systems, 3500 John A. Merritt Blvd., Box 9501, Nashville, TN 37209, USA}
% \offprints{I. Pillitteri}

\email{ipillitteri@cfa.harvard.edu}             

%   \date{Received; accepted}

\begin{abstract}
{\ We report on two XMM-Newton observations of the planetary host star \hd .  
The system has a close in planet and it can potentially affect the coronal structure via 
interactions with the magnetosphere.}
{We have obtained X-ray spectra and light curves from EPIC and RGS on board XMM-Newton
which we have analyzed and interpreted.
We reduced X-ray data from primary transit and secondary eclipse occurred  in April 17th 2007 
and May 18th 2009, respectively.} 
In the April 2007 observation only variability due to weak flares is recognized. 
In 2009 HD189733 exhibited a X-ray flux always larger than in the 2007 observation.
{\ The average flux in 2009 was higher than in 2007 observation by a factor of 45\%.}
{During the 2009 secondary eclipse we observed a softening of the X-ray spectrum  
significant at level of $\sim3\sigma$. 
Further, we observed the most intense flare recorded at either epochs. This flare occurred 
3~ks after the end of the eclipse. {\ The flare decay shows several minor ignitions 
perhaps linked to the main event and hinting for secondary loops that emit triggered by the
main loop. 
Magneto-Hydro-Dynamical (MHD) simulations show that the magnetic interaction between 
planet and star enhances the density and the magnetic field in a region comprised
between the planet and the star because of their relative orbital/rotation motion.}
X-ray observations and model predictions are globally found in agreement, despite 
the quite simple MHD model and the lack of precise estimate of parameters including the alignment 
and the intensity of stellar and planetary magnetic fields.} 
Future observations should confirm or disprove this hypothesis, by determining whether flares 
are systematically recurring in the light curve at the same planetary phase.  
\end{abstract}

   \keywords{stars: activity -- planetary systems -- stars: magnetic fields -- stars: individual (HD189733) }

   \maketitle
%
%________________________________________________________________

\section{Introduction}
Recent work indicates that exoplanets, and especially Jupiter class planets at a distance 
$\la 0.1$ AU from their parent stars (the so called {\em hot Jupiters}), 
are in a unique X-ray environment 
and the effects of stellar X-rays may be significant to their evolution.
The X-rays have been cited as the cause of excess heating of the planet, 
which can induce mass loss \citep{Lammer03}. 
Further, it has been argued that the magnetic fields of the star and the planet can interact  
when the latter is orbiting at a distance of very few stellar radii as in the case of \hd\
\citep{Lanza08,Lanza09,Cohen09}.
The complex magnetosphere that is formed by the star+planet system can drag material
from the outer atmosphere of the planet and funnel it onto the stellar surface. 
The accretion of gas on the star that eventually could arise from this phenomenon
can provoke shocks that heats the plasma up to a few million of degree and thus emits
in X-rays. Furthermore, the enhanced magnetic field near the stellar surface 
can form very active regions on the star increasing the activity of the star
in the X-ray band.
Recent statistical analysis of X-ray activity of stars possessing hot-Jupiters 
indicates that stars with close orbiting planets show enhanced X-ray emission 
\citep{Kashyap08}.  
In the optical band, \citet{Shkolnik03} have detected enhanced activity of Ca H~\&~K in phase
with planet's orbital period in HD~179949.

A typical ``hot Jupiter" receives $\sim2\times10^4$ times more radiation from 
its star than Jupiter does from the Sun, and orbits at a distance of
only $\sim$10 stellar radii.  Unlike solar system planets, where
residual heat from formation and gravitational settling still plays an important part 
in the energy budgets of Jupiter and other giant planets, the energy budget for hot
Jupiters is completely dominated by the strong radiation that they receive from
their parent stars.
This energy budget is poorly constrained in the X-ray and UV regimes
by standard stellar atmosphere models.  Thus,
understanding the nature of this input and the planet-star
interactions is essential for understanding the properties and the
evolution of these planets. 
As pointed by \citet{Lammer03}, the stellar X-ray and UV fluxes in 
their irradiance calculations strongly determine the amount of losses of gas 
from the outer atmospheres of close-in exoplanets, otherwise the losses of volatiles in 
models neglecting UV/ X-ray irradiation are proved to be unrealistically low. 
The UV/X-ray flux from the star plays a crucial role in
determining the photochemistry of the planet's upper atmosphere. The photo-chemical 
products can act, in turn, as high-altitude absorbers, create thermal inversions, and otherwise alter 
the observable properties of the planet's upper atmosphere. These effects are important when  
observing at lower energies/longer wavelengths (\citealp{Burrows08}, \citealp{Liang04})

\hd\ (R.A. $20^h00^m43.7^s$, Dec. $+22^d42^m39.1^s$) is a K~1.5V type star at a 
distance of 19.3~pc,  of mass 0.81 M$\odot$, radius 
0.76 R$\odot$ and a rotational period of $\sim11.95d$ \citep{Henry08}. 
It is in a wide binary system and \hd B, the other stellar member of the binary system,
is a M4V type star (with range: M3.5--M5, $M \sim 0.2 M_\odot$, \citealp{Bakos2006}) orbiting 
at a mean distance of $\sim 220$ AU (12.3$\arcsec$ apparent separation) 
with a period of 3200~yr in a plane nearly perpendicular to the Earth-\hd A line of 
sight. % \citep{Bakos2006}.
It is thought that the age of this system is greater than or equal to 0.6 Gyr \citep{Melo06}.
{\ The magnetic field is estimated of a few tens of Gauss, a toroidal component
and higher multipole components are presumed and a strong shear of field lines due to differential 
rotation is inferred \citep{Moutou07}.}

\hd A hosts a Jupiter size planet (\hd b, $M \sim 1.15$ $M_{Jup}$, $R \sim 1.26 R_{jup}$ ) 
at a distance of only 0.031 AU (8.75 stellar radii) and with an orbital period of 2.219d
\citep{Bouchy05}, much lower than the stellar rotational period. 
\hd b is thus classified as a {\em hot Jupiter} planet type.
The orbit is nearly circular and orbital plane is parallel within a few degrees of the 
Earth-primary line of sight. This system has been the subject of many studies since its discovery, 
given its close distance and favorable geometry.
Given Q-values for giant planets in the solar system, it is thought that orbital radii 
of a few hundredths of an AU from parent star reduces hot Jupiters tidal dissipation 
timescales to a few Myr, and tidally locked hot Jupiters should be the norm.
Under this hypothesis, \hd b should show a pronounced temperature difference between the
day-side and the night-side hemispheres. Instead  \citet{Knutson07} found only 
a modest difference of  temperature between day-side and night-side hemispheres, 
suggesting that the thermal energy transport and thermalization in the planet's atmosphere 
is quite efficient. The same study shows that a bright spot is present in the planet surface 
displaced by $16 \pm 6^o$ east longitude with respect to the sub-stellar point.

Observations with ACS camera on~board {\em Hubble} allowed \citet{Pont07} to infer that planet
transits cover spots on the stellar surface with sizes of order of 80000 km. \hd A is known
to be an active and variable K type star with strong magnetic activity, as inferred from CaII H\&K  
lines \citep{Moutou07}. 
Through modeling the observed Rossiter-McLaughlin effect (the apparent Doppler shift occurring
for the transit of the planet at the stellar limbs) \citet{Winn06} find that the spin stellar 
axis and the orbit normal are misaligned of only 1.4$^o$, hinting that migration of the planets
does not change this alignment.

{\ Due to its close distance and the favorable orientation of the planetary orbit plane, 
\hd\ is an ideal laboratory for the study of close in planets. Among the plethora of known exoplanets, 
the details on the atmosphere of \hd b obtained through optical and IR spectroscopy are unique.}

In this paper we report on two observations of \hd\ system obtained in X-ray band with 
{\em XMM-Newton} satellite at the time of a planetary transit and a secondary eclipse, 
respectively.  
X-ray observations can constrain  model predictions, 
reduce degeneracy in their parameter space and help MHD simulations to 
infer physical and geometrical properties of the complex magnetosphere created by star 
and planet.

The structure of the paper is as following: 
in Sect. \ref{data} we describe the observations and the data analysis; 
In Sect. \ref{results} and \ref{discussion}  we show the results and discuss
them and in Sect. \ref{conclusions} we present our conclusions.

%__________________________________________________________________
\begin{figure}
\includegraphics[width=\columnwidth]{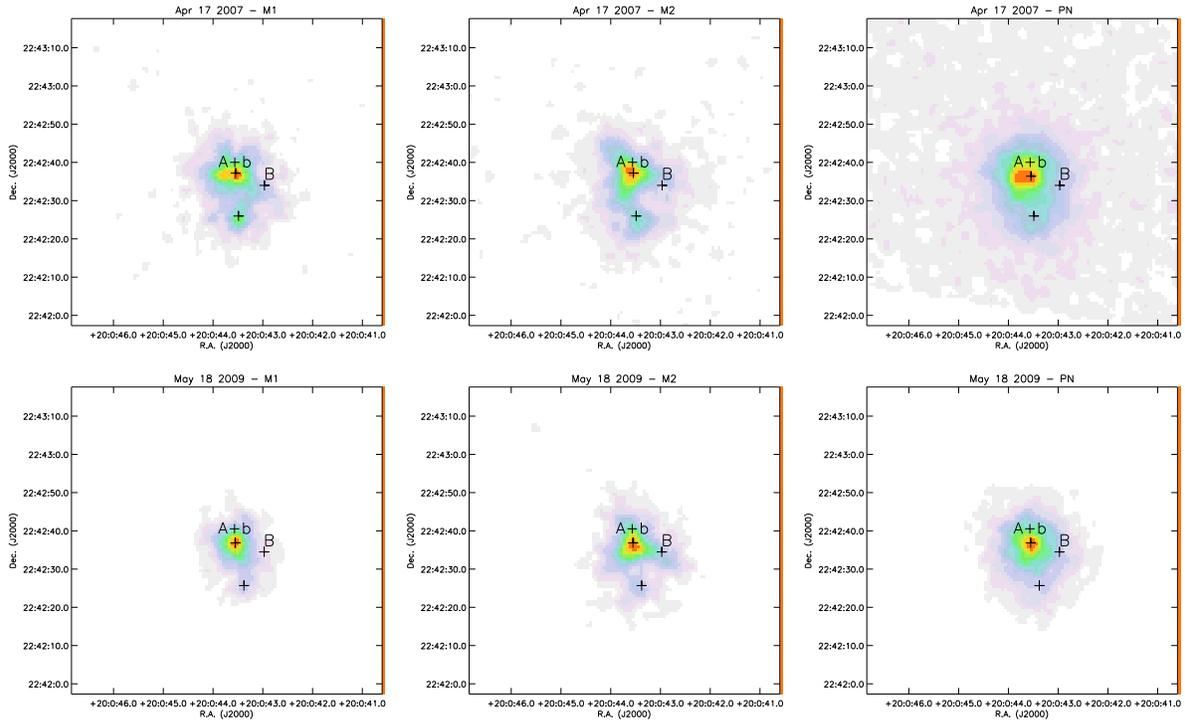}
\caption{\label{img}EPIC images of \hd A+B observed with XMM-Newton in 2007 April (top row)
and 2009 May (bottom row). Images of MOS 1, MOS 2 and PN are shown in sequence from left to right.
{\ The companion \hd B is not visible in both observations. A bright source with uncertain 
optical counterpart is visible to the south of \hd A, and it has faded in 2009.}}
\end{figure}

\begin{figure}
\begin{center}
\includegraphics[width=7cm]{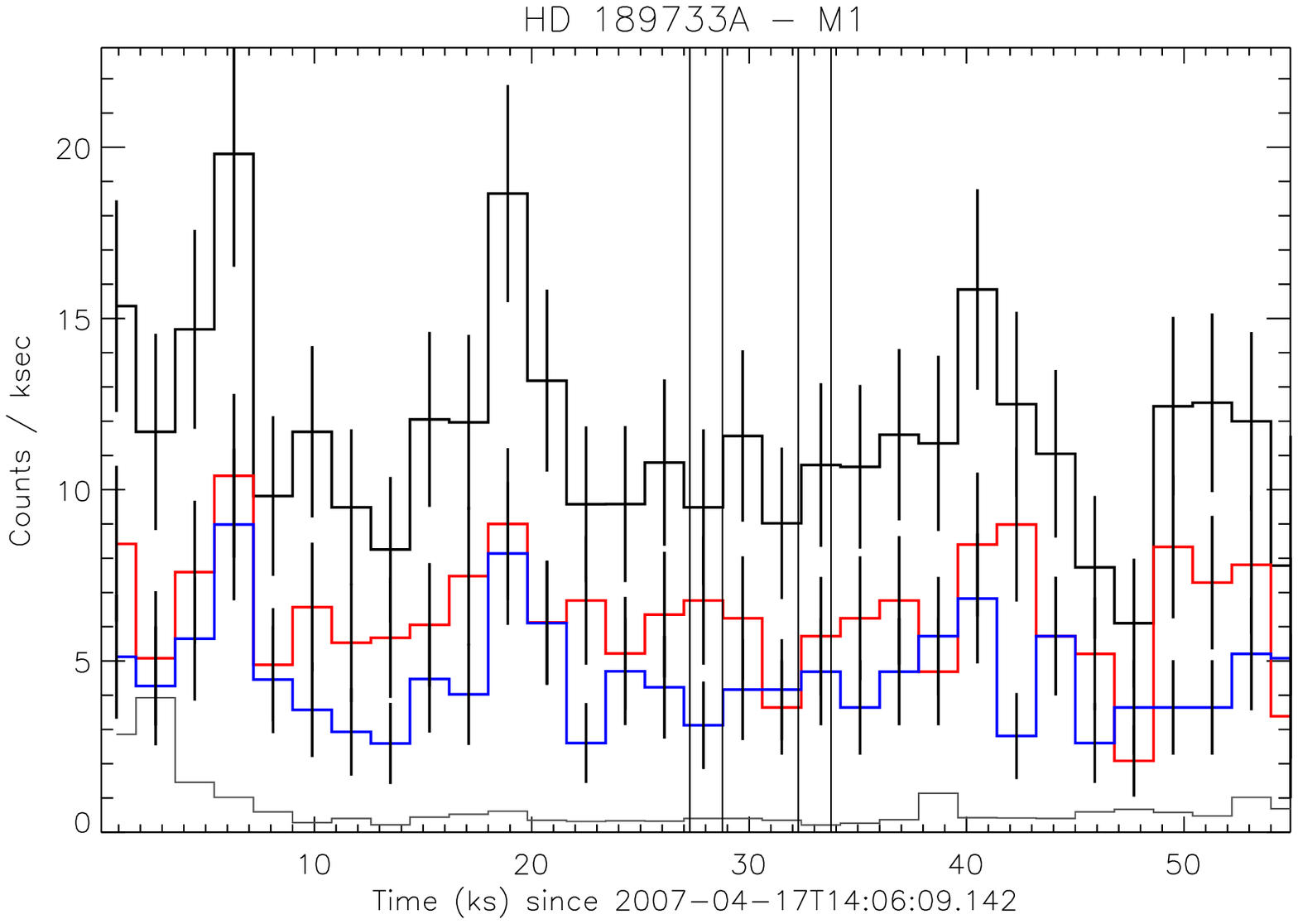}
\includegraphics[width=7cm]{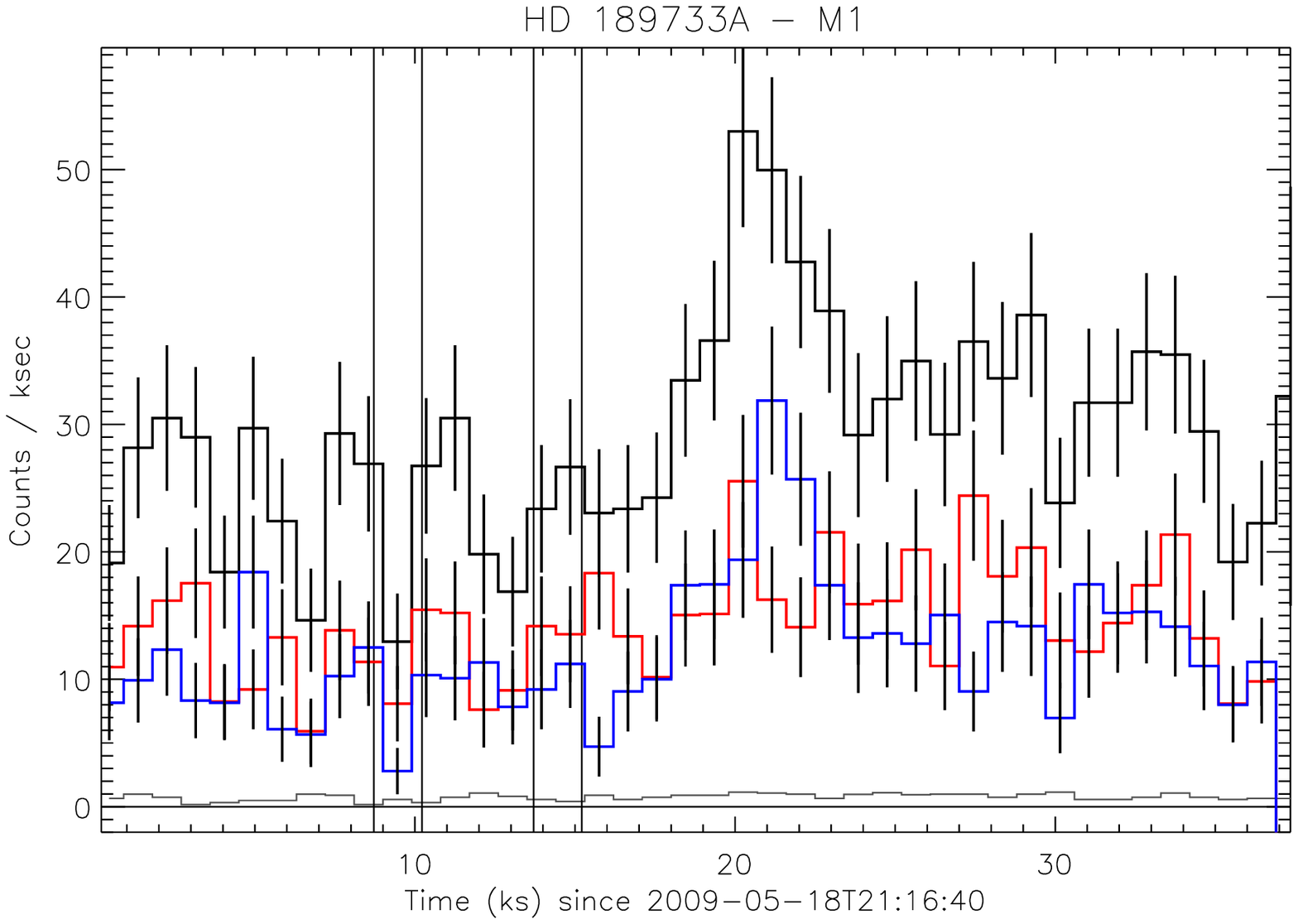}
\includegraphics[width=7cm]{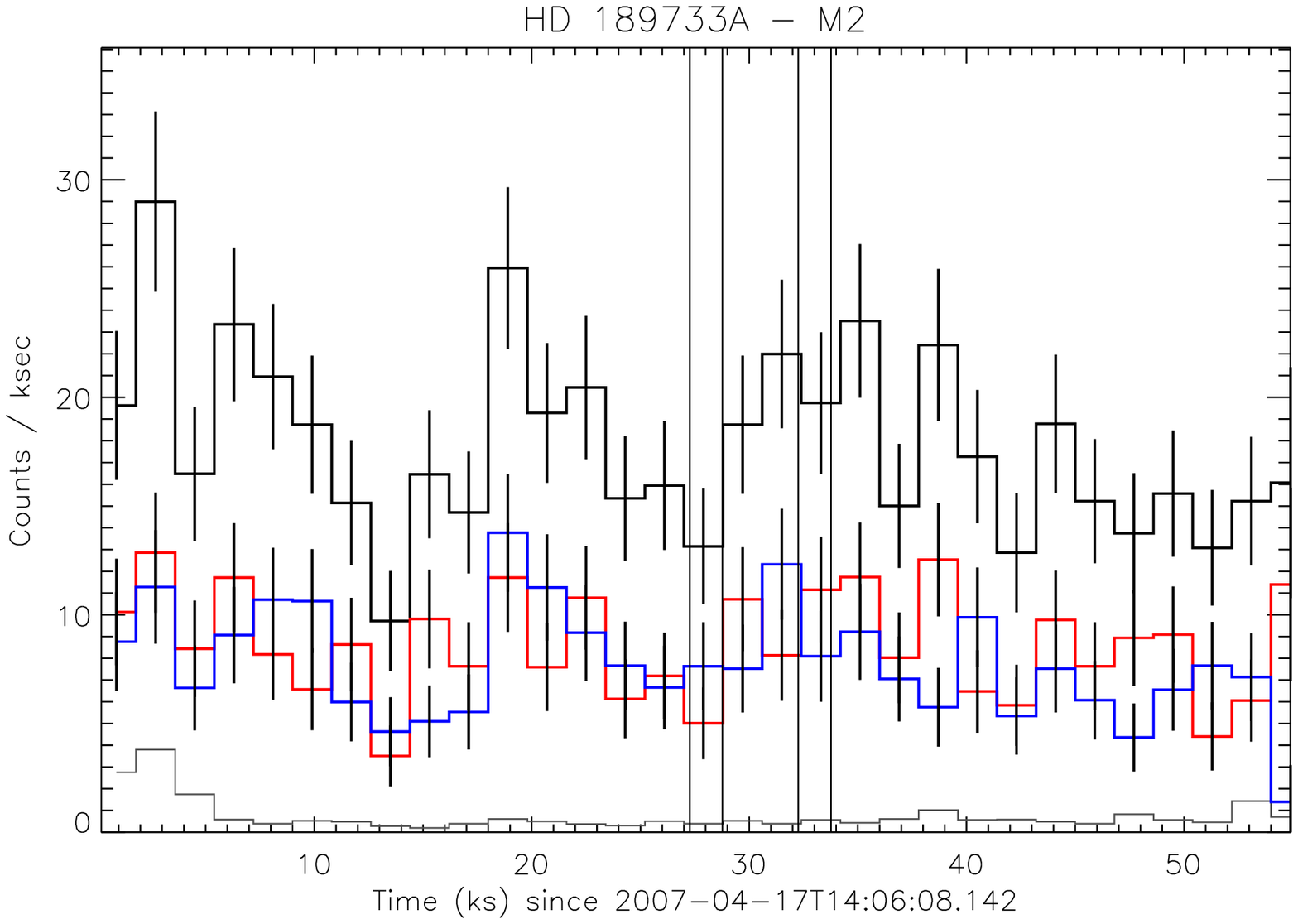}
\includegraphics[width=7cm]{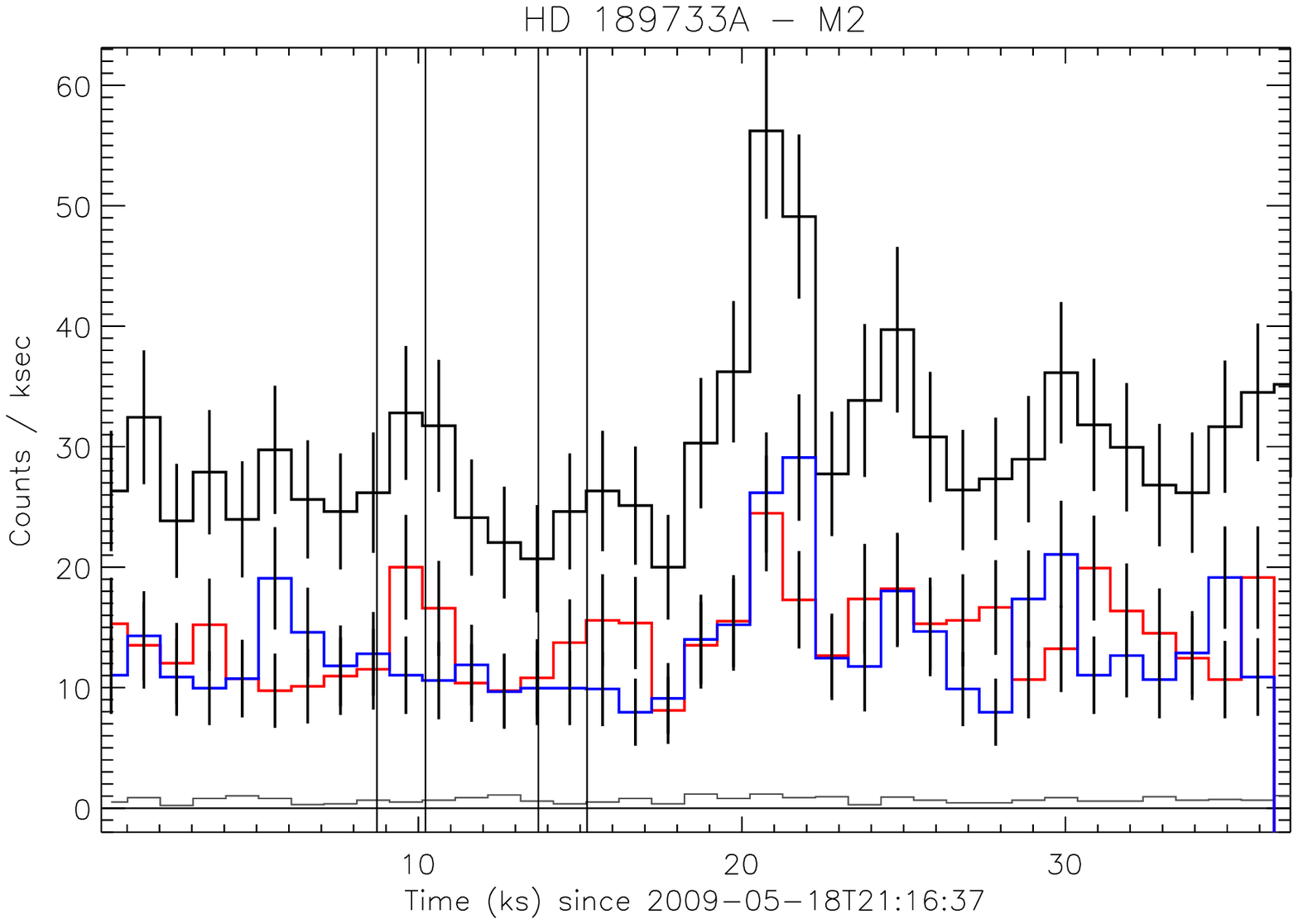}
\includegraphics[width=7cm]{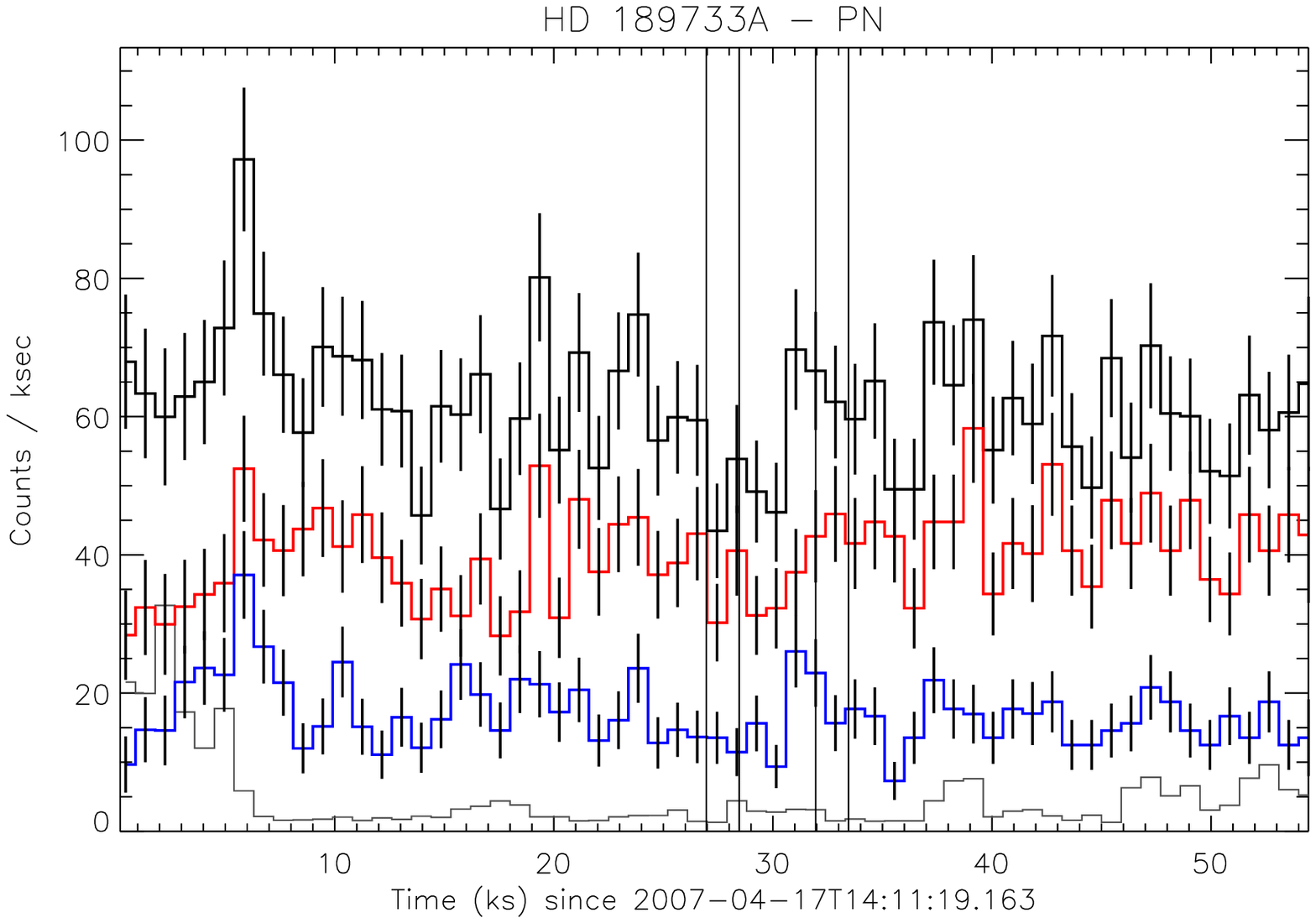}
\includegraphics[width=7cm]{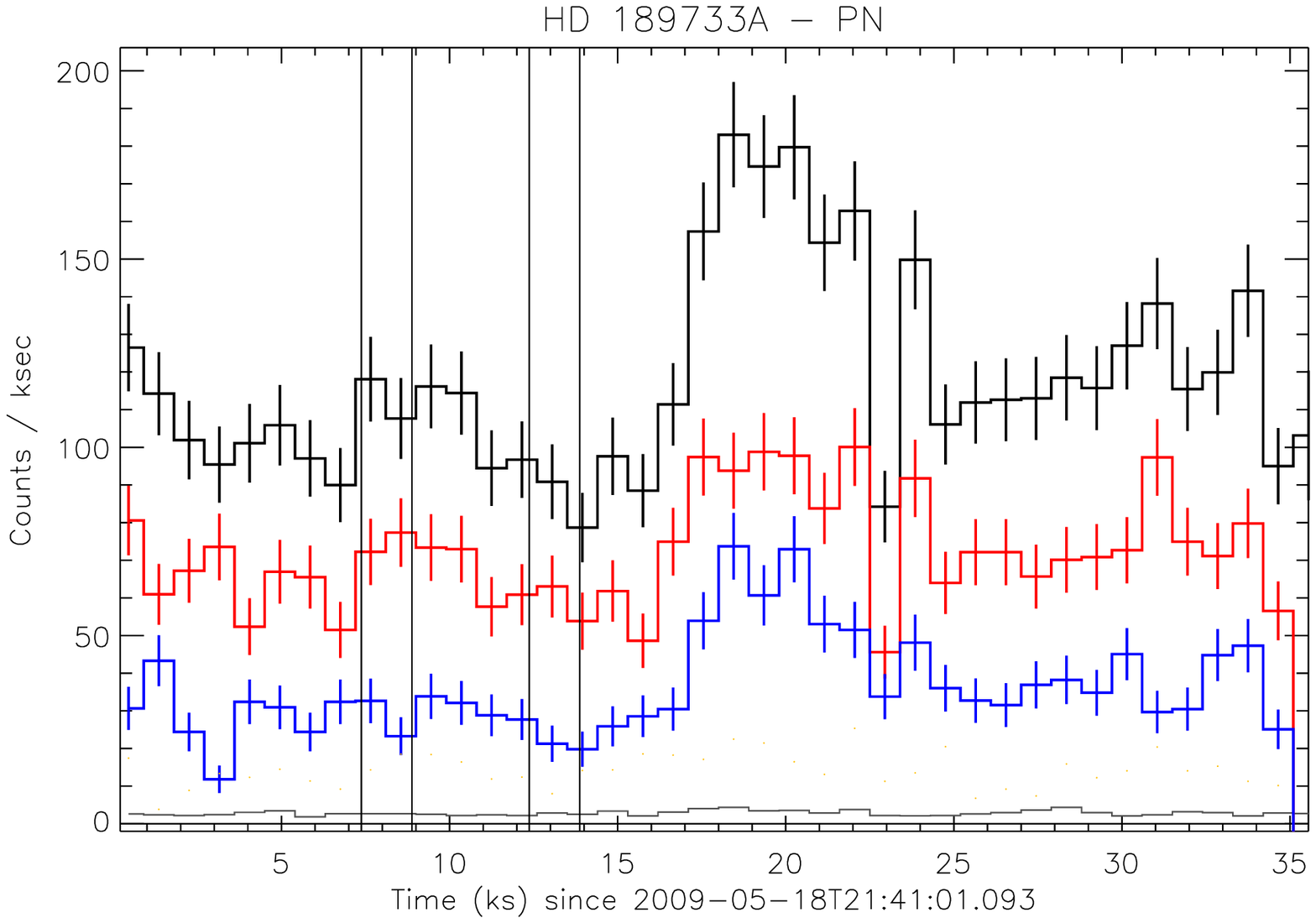}
\end{center}
\caption{\label{lc} Light curves of \hd . 
On the left column: light curves of MOS1, MOS2 and PN acquired during the 2007 observation.
On the right column the same for the 2009 observation. Error bars are at $1\sigma$ level.
The red light curve is in the 0.3-0.8 keV energy band 
while the blue light curve is in 0.8-2.5 keV energy band. Gray curve is the background.
Bin size is 900 s for PN and 1800 s for MOS 1 and 2. 
}
\end{figure}

\section{\label{data} Observations and Data Analysis}
We have obtained an observation  of \hd\  with {\em XMM-Newton} during the 
eclipse of the planet (secondary transit, when the planet passes behind the star 
as seen from the Earth) that occurred on $18-19$ May 2009  (Obs \# 0600970201).
We have also analyzed a previous {\em XMM-Newton} archival observation of \hd\ 
during the planetary transit of 17 April  2007 (Obs. \# 0506070201, P.I. P. Wheatley). 
{\ A source close to \hd\  is visible in both observations, especially in 2007,
while the M-type companion \hd B  is undetected. Obviously, the planet is not resolved.
Fig. \ref{img} shows the EPIC images of \hd\ obtained in 2007 (top row) and 2009 (bottom row), 
respectively.}
The XMM {\em Observation Data Files} of both observations were processed with SAS (ver. 8.0)
to obtain event tables of MOS1, MOS2 and PN calibrated both in energy and position.
We restricted the analysis to events in the $0.3-8.0$ keV energy band and to events that 
triggered at most two adjacent pixels as suggested by the SAS guide.
We have extracted events of \hd A from a circular region that avoids 
the contamination from the M star companion \hd B. 
From the events selected in this region we have obtained light curves (Fig. \ref{lc}) 
and EPIC spectral of \hd A. 
For the analysis of light curves and spectra we focus primarily on the PN data in this work 
because of its higher efficiency in collecting counts with respect to both MOS cameras
(4557 PN counts vs 1166 MOS1 and 1190 MOS2 counts in the source extraction region, respectively).
A likely loss of data telemetry causes a dip at $\sim23$~ks with a duration of $\sim900$~s 
in the light curves of 2009 observation, more marked in PN data.

\begin{figure}
\includegraphics[width=0.6\columnwidth,angle=0]{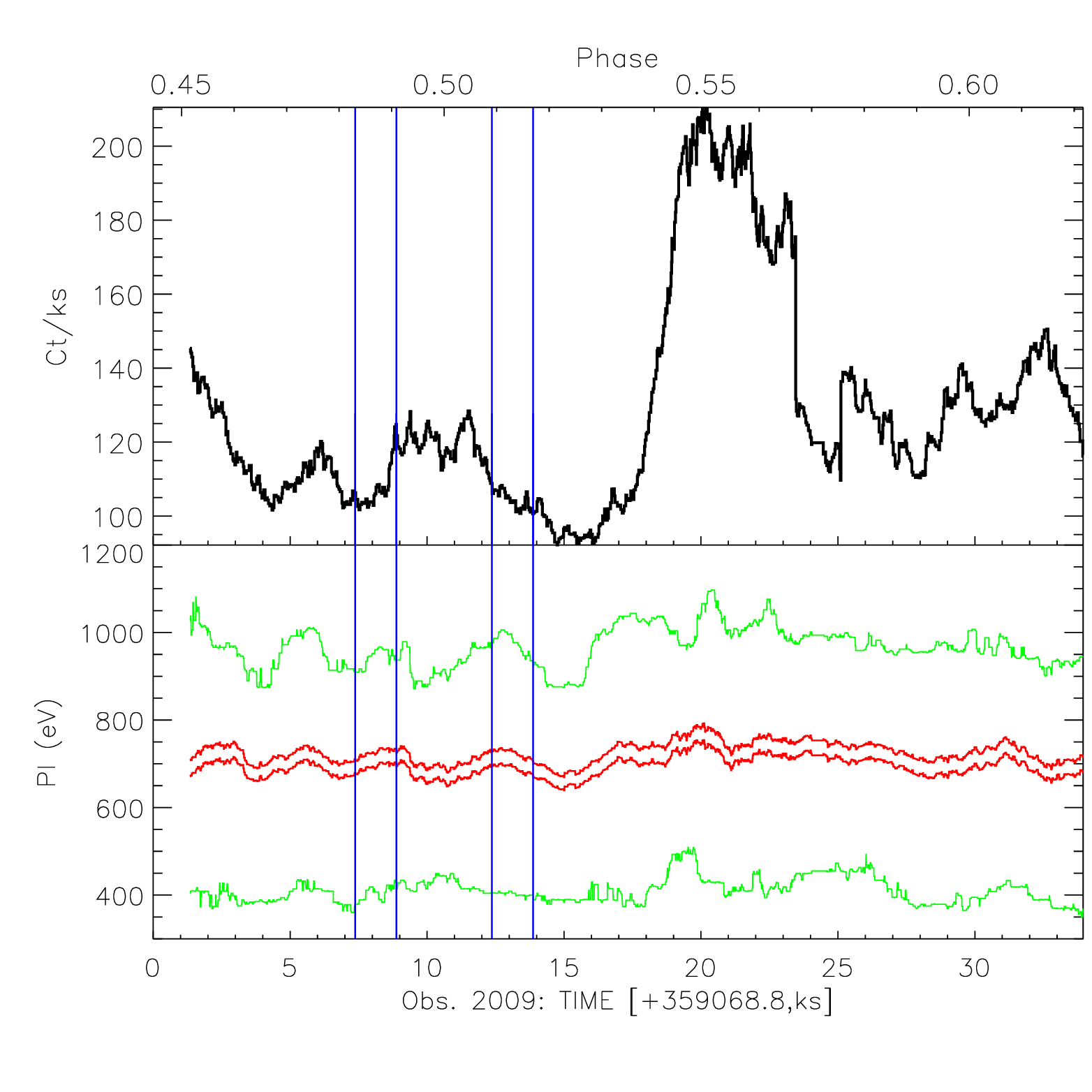}
\includegraphics[width=0.4\columnwidth,angle=0]{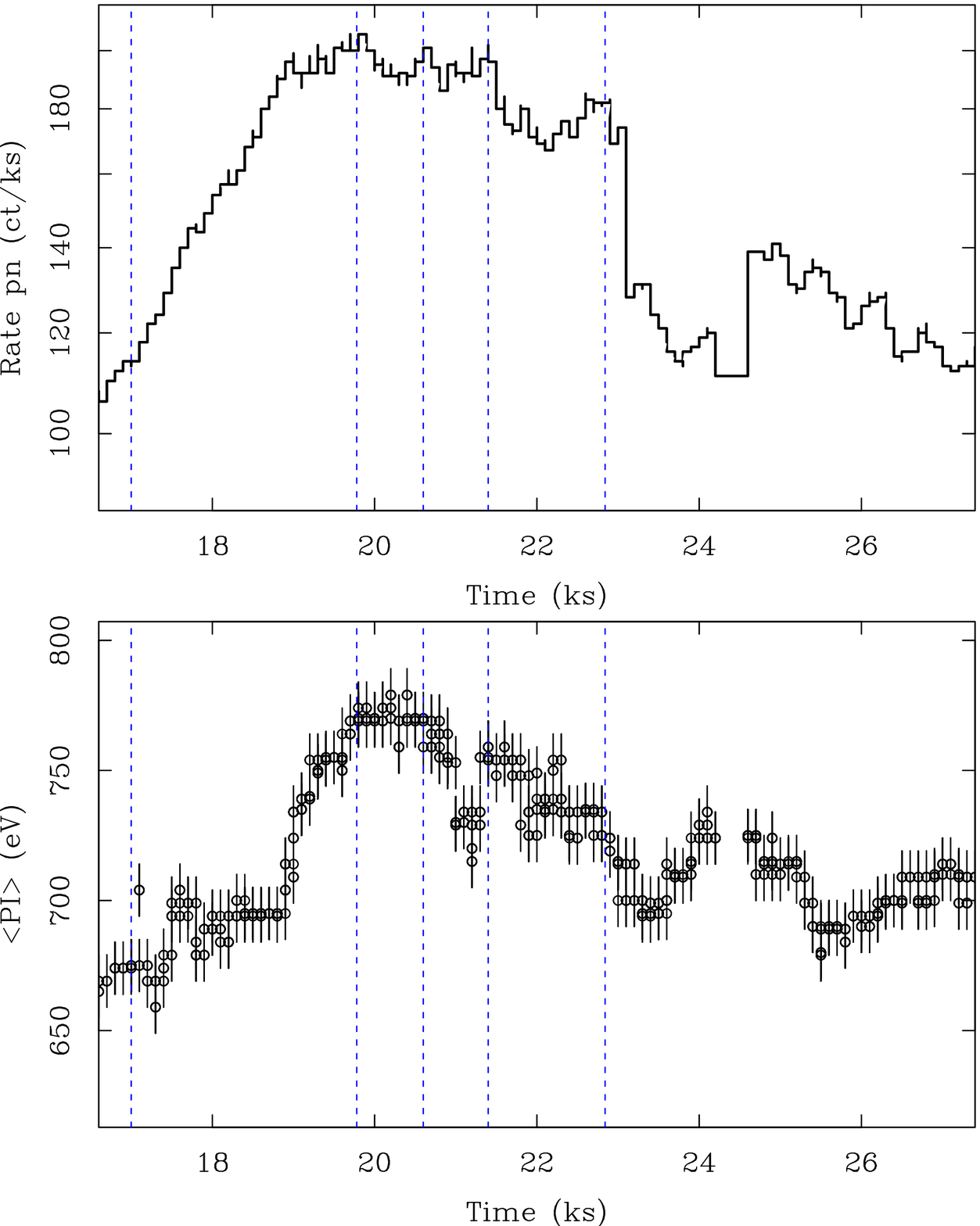}
\caption{\label{smoothed} 
Intensity (top panels) and spectral variation (bottom panels) of \hd A
during 2009 (secondary eclipse) registered from PN.  
On the left panels the curves relative to the whole exposure time.
Vertical lines denote the 1st, 2nd, 3rd and 4th contact. 
In the bottom left panel, the green curves represent  the 16\% and 84\% 
quantiles of the distribution of a sample of 200 photons, and show the overall 
width of the PI spectral distribution.  
Red curves represent the mean $\pm1 \sigma$ standard deviation of the mean of the
sample of PI between 0.3 and 1.5~keV, and show the temporal variation of the spectrum.
On the right panels we plot the rate and the mean of PI during the flare with a log
scale in the y-axis of rate. 
Vertical lines denote the beginning of the flare, the peak, and three secondary impulses.}
\end{figure}

In order to characterize the variations of the source at short timescales,
we consider the pulse invariant (PI) spectrum of PN events in a moving window comprised of
200 events.  We compute the average energy of the spectral distribution,
as well as the median, the standard deviation, and the 16\% and 84\%
quantiles for a contiguous set of 200 photons (see Figure \ref{smoothed}).  The window
is moved by adding five events at the end and removing five from the beginning.
We obtain a smoothed light curve in the same manner, with the time interval
between the first and last events in the window suitably corrected for the
presence of gaps in the exposure time.  A background is estimated
for each interval and is subtracted from the light curve\footnote{Note that background
is ignored for the spectral distribution both due to the relative constancy
of the background spectrum and the numerically small correction to the
spectrum; see Figure \ref{lc}}.

We have also extracted RGS spectra to have a higher resolution 
spectrum ($\delta \lambda \sim 0.04\AA$) in order to observe the behavior of
features, especially O~VII triplet and its changes during the observation. 
The ratio of forbidden to recombination lines from OVII triplet offers a diagnostic of 
the plasma density. For the main flare observed in the light curve of 2009 observation,
we have performed time resolved spectroscopy of PN data 
by considering the spectrum during the flare,  as well as before and after the flare.
The spectra of EPIC PN have been analyzed with XSPEC software to obtain best fit parameters
from thermal models (APEC) with one or two temperatures. 
Given that absorption is very low due to the short distance to the star, we have considered
only unabsorbed spectral models. 
The free parameters were the temperatures, the global abundance and the emission measures.

\begin{figure}
\includegraphics[width=\columnwidth,angle=0]{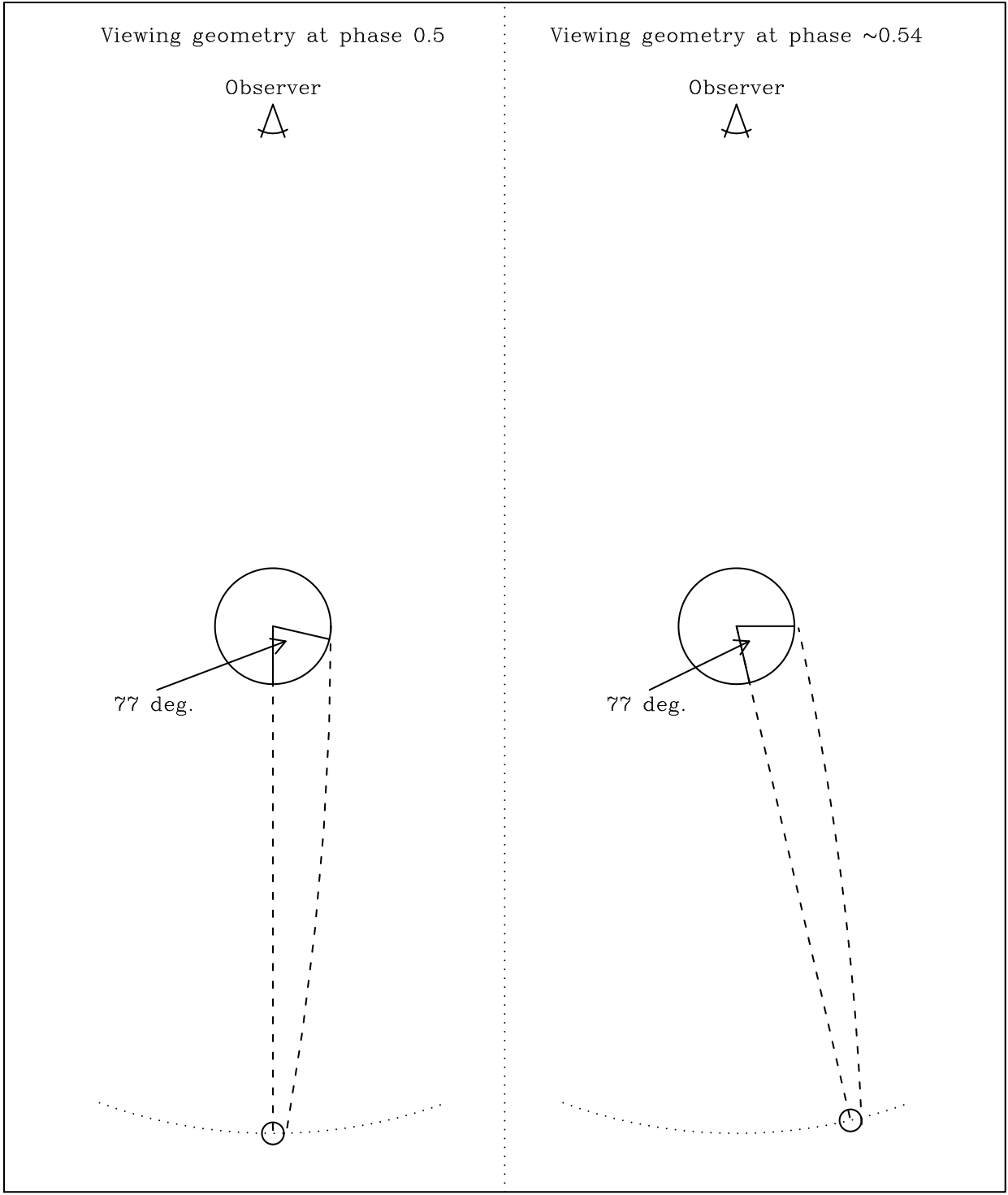}
\caption{\label{geometry} Scheme of the angle between sub-planetary point and active region
where the large flare could have occurred in 2009.}
\end{figure}

\begin{figure}
\includegraphics[width=\columnwidth,angle=0]{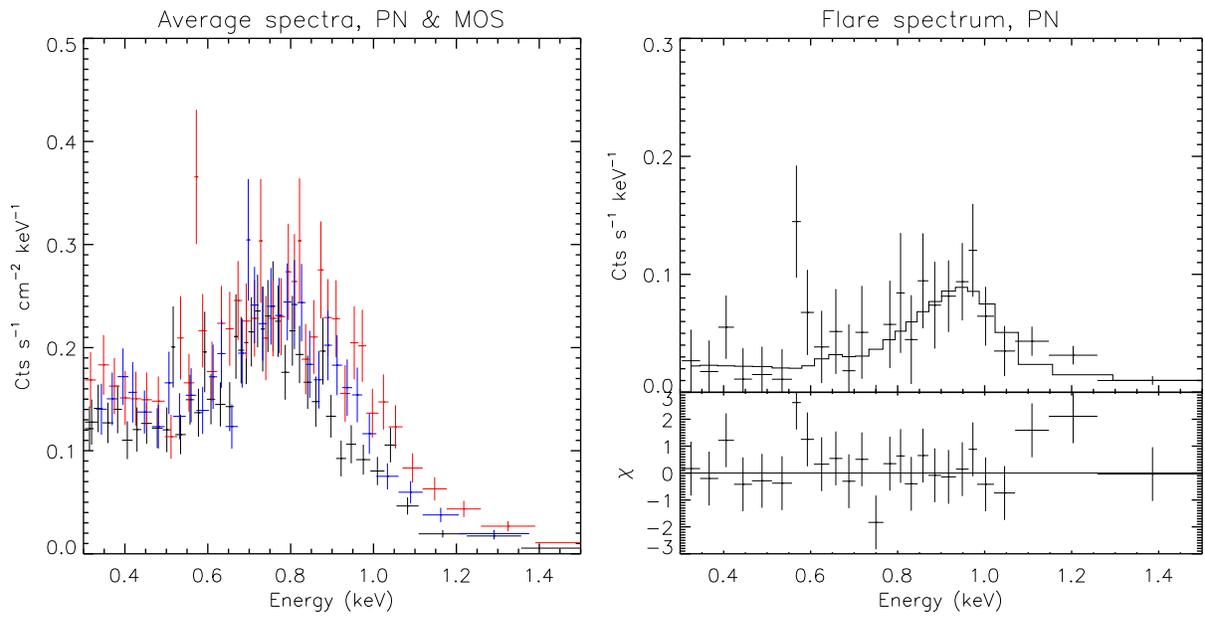}
\caption{\label{sp} Left panel: PN spectra of \hd\ A during the three phases, 
pre-flare (0--15~ks, black), flare (15--25~ks, red) and post-flare (25--35~ks, blue).
Right panel: flare spectrum (minus pre-flare spectrum) with best fit model and residuals.
The region of OVII at $\sim 0.57$ keV ($\sim 22$\AA ) is enhanced during the flare.}
\end{figure}

\begin{figure}
\includegraphics[width=\columnwidth,angle=0]{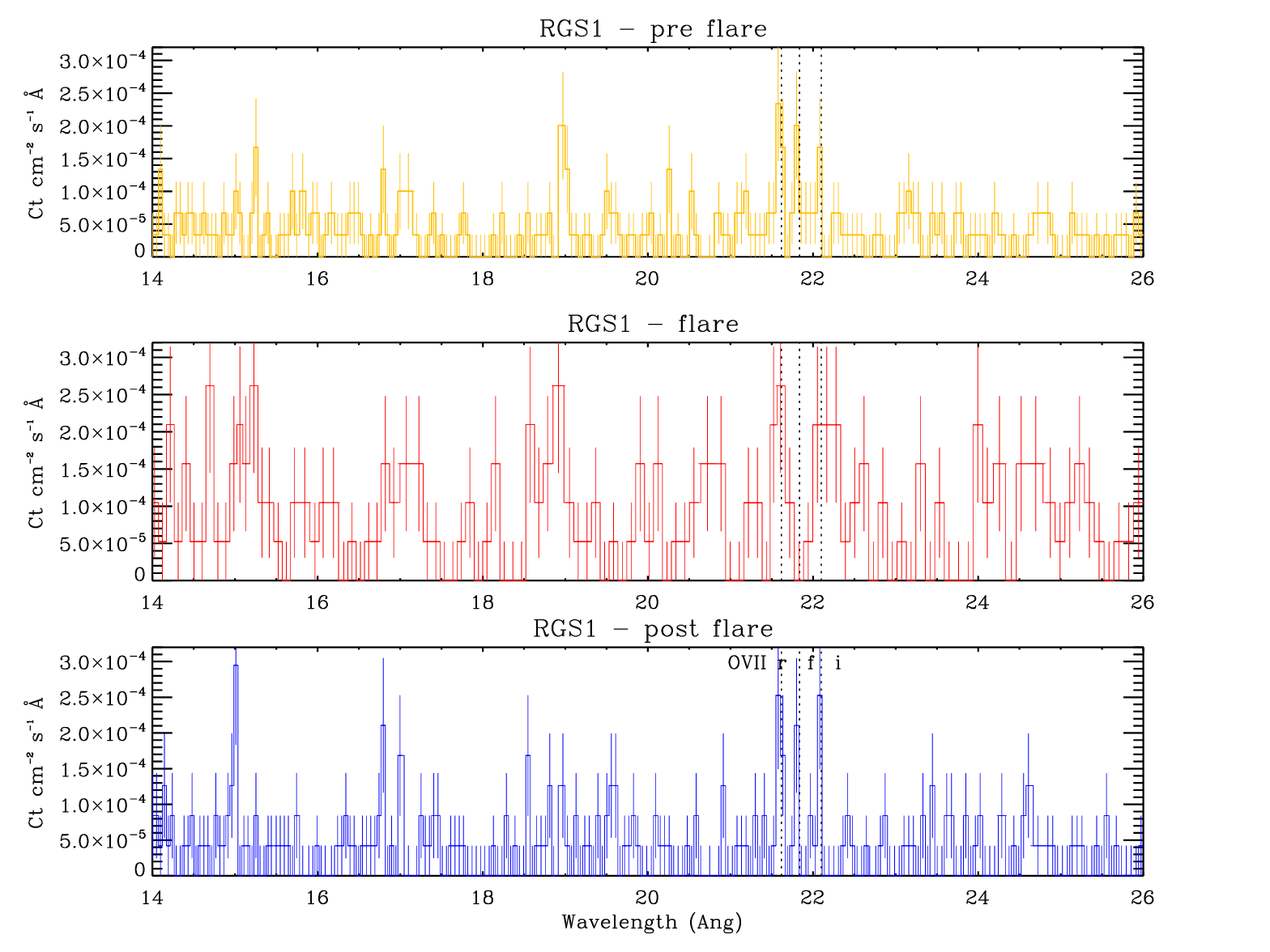}
\caption{\label{sprgs} RGS~1 spectra of \hd A during the three phases:
pre-flare (0-15~ks), flare (15-25~ks) and post-flare (25-35~ks).}
\end{figure}

%______________________________________________________________
\section{\label{results} Results}
In Fig. \ref{lc} we plot the light curves from the 2007 and 2009 observations
in three bands:  0.3-0.8 keV (red), 0.3-2.5 keV (blue) and 0.3-8.0 keV (black).
The background curve is shown in gray.
The duration of the planetary transit and the secondary eclipse is 
$\sim$108~min from 1st to 4th contact, i.e. 6480~s. 

The 2007 light curve shows a modest flare in coincidence of the planetary transit as 
the spectrum seems to harden at $\sim$30~ks ($\phi \sim 0$) and decays 
exponentially post-transit. 
We find that a 1-T thermal model with plasma temperature of $\sim0.6$ keV reasonably 
fits the spectrum during the first flare. This is the most prominent feature in the 2007 
light curve. 
The remaining part of the observation is best modeled with a 2-T thermal model with a cool 
component ($kT= 0.24$ $keV$) and a hot component ($kT=0.56$ $keV$) very similar to the first 
flare temperature.
The unabsorbed flux in the 0.3-8 keV band energy is $\sim2.5 \times 10^{-13}$ erg s$^{-1}$ cm$^{-2}$
for a luminosity of $L_\mathrm X = 1.1\times 10^{28}$ erg s$^{-1}$.
Table \ref{spec} shows the results of the fits. Abundances are found at $Z \sim 0.15-0.2 Z_\odot$, 
as in active stellar coronae and very young stars \citep{Maggio07}.

During the 2009 observation the star had a X-ray activity more pronounced than in 2007.
The mean energy of the 2009 spectrum is $\sim700$ keV and the PN rate is on average 
$\sim65$ ct ks$^{-1}$ during most of the observation, reaching a peak of $\sim95$ ct ks$^{-1}$ 
during the first flare. In 2009 the PN rate of \hd A is always larger 
than in 2007: the lowest rate is in fact $\sim95$ ct ks$^{-1}$ in 2009 with a peak of 
$\sim350$ ct ks$^{-1}$ (see Fig. \ref{smoothed}).

During the secondary eclipse, the light curve shows a {\em softening} of the spectrum. 
From Fig. \ref{smoothed} we see that the mean of the PN distribution of PI gets softer
during the eclipse.  The mean photon energy in the $0.3-1.5$~keV range before and after 
the eclipse is $\sim700\pm10$ ($1\sigma$), while during the middle of the eclipse it 
is $\sim660\pm10$~eV ($1\sigma$), thus it is significant at a level 
of $3\sim\sigma$. 
During the eclipse the rate increases. The quartile curves show that the soft part of the 
spectrum increases while the hard part of the spectrum decreases. 

The light curves in Fig. \ref{lc} and \ref{smoothed}
show a flare with a decay of $\sim10$ ks that occurs 3~ks
after the end of the planetary eclipse. The flare decay is characterized by 
several smaller impulses after the main ignition\footnote{A likely loss of telemetry is causing
the sharp dip in the decay around 23~ks.}.
The flare phase is $\phi\sim0.54$ which would correspond to the planet being $16^o$ past the
earth--star line of sight (see Fig. \ref{geometry}).
In the soft band (0.3--0.8 keV) the light curve of PN has a step-like profile (Fig. \ref{lc}), 
characterized by an almost flat high level whereas in MOS 1 and 2 the soft band light 
curves show a more usual peaked profile. 
The difference of the flare profile in PN and MOS is very likely due to 
their different responses at low energies {\ with the PN being more sensitive to low 
energies than MOS.}\footnote{ The ratio
of normalized PN and MOS effective areas shows that PN is proportionally 
more sensitive than MOS by a factor $\sim2$ in the range $0.3 - 0.7$ keV.}
{\ The plasma spectrum becomes hard at the flare peak as shown by the median
and quartile curves (right panel in Fig. \ref{smoothed}). We deduce that the temperature of the 
plasma increases approaching the peak 
}

To investigate the characteristics of the flare we have analyzed the PN spectra in the
intervals $0-15$~ks, $15-25$~ks, and $25-35$~ks (chosen by visual inspection of the light curve). 
In Table \ref{spec} we report the best fit parameters and 90\% confidence ranges 
of the model fit to the spectra in the three phases.
We also list the result of fitting the flare spectrum subtracted from the pre-flare spectrum.

The flare spectrum in 2009  shows a modest increase in temperature,
changing from $\sim0.4$ keV prior to the event to $\sim0.9$ keV (flare spectrum minus pre-flare 
spectrum). 
The temperature of the flare is similar to that observed in solar flares ($\sim 10$ MK).
The total flux of \hd A+b increases by a factor of 85\% and the  emission measure by 11\%. 
The flux in the 0.3--8 keV band due to the flare is $1.3\times10^{-13}$ erg~s$^{-1}$~cm$^{-2}$.
Some features of the coronal spectrum are enhanced during the 
brightening. Fig. \ref{sp} shows the PN spectra in the three phases: black is pre-flare, red is the
flare and blue is after the flare. The most notable difference is around 0.57 keV or 22$\AA$
where the O~VII triplet is located (recombination: $r = 21.62\AA$, intercombination: $i=21.84\AA$, 
forbidden: $f=22.10\AA$). This region of the spectrum remains in excess at a level of 
$\sim2\sigma$ with respect to the best fit thermal model.

We explore this spectral region in details by means of the RGS~1 spectrum that allows a much higher 
resolution with respect to EPIC cameras in order to distinguish between the $r$, $i$, and $f$ 
triplet lines of O~VII. {\ The relative intensity of these lines offers a diagnostic 
of the plasma density, wherein the forbidden line is formed when the plasma has low density and
the corresponding energetic level is de-populated by collisions under increasing plasma density.}
Fig. \ref{sprgs} shows the spectra  in the three phases as above. 
We have rebinned the spectra, by decreasing the instrumental resolution by a factor 5 to
enhance the signal. In these spectra we can easily distinguish between $r$, $i$ and $f$ lines. 
During the flare the $i$ line seems to disappear, and the $f$ line is less luminous.
After the flare the $f$ line remains quite luminous and this fact suggests that low density plasma 
continues to be visible after the heating. 
There are indications that the $r$ recombination line is shifted by $\gtrsim{150}$ km/s, 
which is consistent with the orbital speed of the planet ($\sim$150 km/s).  
\citet{Linsky2010} looked for UV signatures of ionized gas co-orbiting with HD 209458b, 
while they found evidence of an exosphere, they concluded that higher S/N were needed to ascertain the 
velocity structure. Likewise, our data are not sufficient for such detection.
Other visible changes during the flare which may be related to the increase of temperature 
are: an enhancement of O~VIII line at 18.96\AA\, Fe~XVII lines at 15\AA\ and 17\AA\ and an 
enhancement of the continuum level, 

In the framework of modeling of the flares in the stars in analogy with solar flares, several studies
have been done involving both simple models based on scaling laws and more detailed 
MHD simulations \citep[see][and references therein]{Reale07}. 
A simple model for the flare is given by a loop formed by the magnetic field in which the 
plasma is constrained to move only along the field lines. 
The heating of the flare is provided at some point of the loop, which usually has an aspect 
ratio (base radius on loop semi-length) of $r/L=0.1$ similar to loops observed on the Sun.
If the flare we observed took place in a coronal loop anchored to the star surface, 
a rough estimate of the semi-length of the loop is provided by the formula given by
\citet{Serio1991}. We obtain a semi-length of order of $\sim75$\% R$_\star$ 
or $4\times10^{10}$ cm \citep{Serio1991,Reale1997}, assuming a peak temperature of 0.86 keV 
and no further heating mechanism at work  after the initial loop heating. 
Given that we cannot estimate the true peak temperature, the size we would have derived is a
lower limit. 
We can suppose that the size of the flaring loop is of the order of the stellar radius.
Loops of the order of the stellar radius are inferred also from a detailed modeling of a large flare 
observed in the very active dM star Proxima Centauri \citep{Reale2004}. 
Given the proximity of the planet (at $d \sim 8.75$ R$_{\star}$)
the perturbation on the stellar magnetic field due to the planet could affect
the formation of loops and their evolution.
The unusual flaring volume suggests that the site of the flare should be an extended
region of the corona or an arcade formed of similar loops. 
From the estimates of the emission measure and the volume we derive a
mean electron density in the plasma of $n_e \sim 9\times10^9$ cm$^{-3}$.  

From the rotational period ($P=11.95d$, \citealp{Henry08}) and stellar radius ($R=0.76R_\star$, 
\citealp{Pont07}) we infer a mean rotational velocity of $v_\mathrm{rot} = 3.4$ km s$^{-1}$.
Early studies of MS solar type stars have shown that the X-ray luminosity is well correlated
with rotational velocity \citep{Pallavicini81}, 
given the link between rotation, dynamo efficiency, coronal
magnetic field, X-ray activity and $L_\mathrm X$. Comparing with the Sun we see that 
$L_\mathrm X$ and $v_\mathrm{rot}$ scale 
reasonably well with the  $L_\mathrm X \sim (v\mathrm{sin} i)^2$ law in the non saturated regime of 
L$_\mathrm X$. From the rotational velocity ratio between \hd A and the Sun (2 km s$^{-1}$) 
a mean magnetic field strength $\sim$1.7 times the mean solar field is inferred. 

{\ The M-type companion is not detected in either observations although it could have been 
resolved given the separation of $\sim12\arcsec$. We have estimated an upper limit to its X-ray
luminosity in 0.3-8.0 keV band of $L_\mathrm X \le 9\times 10^{26}$ erg s$^{-1}$.
The non-detection of this star has some implications for the estimate of the age of
the system as we will discuss in sect. \ref{discussion}.}

%______________________________________________________________
\subsection{The X-ray Source near \hd}
\begin{figure}
\begin{center}
\includegraphics[width=\columnwidth]{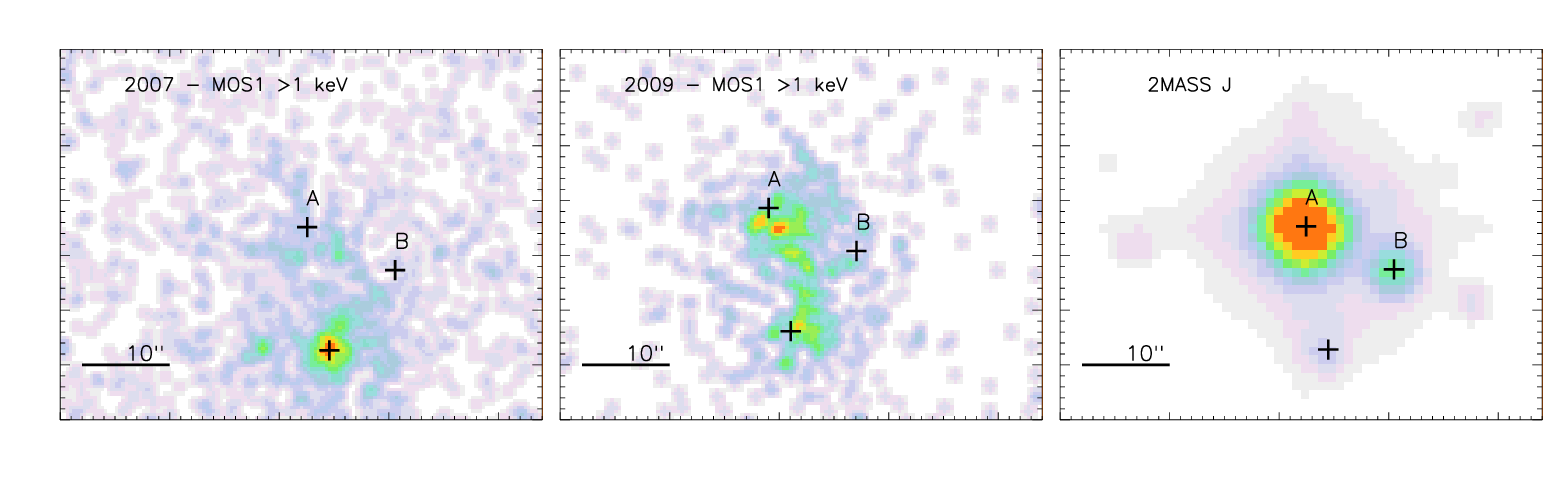}
\caption{\label{isx} EPIC and 2MASS images of the X-ray source at $\sim 12\arcsec$ from \hd .
From left to right: MOS1 $> 1$keV in 2007, MOS 1 $>1$ keV in 2009, 2MASS J-band image.
Positions of \hd A, \hd B and the source are indicated by crosses.}
\end{center}
\end{figure}

\begin{figure}
\begin{center}
\includegraphics[width=\columnwidth]{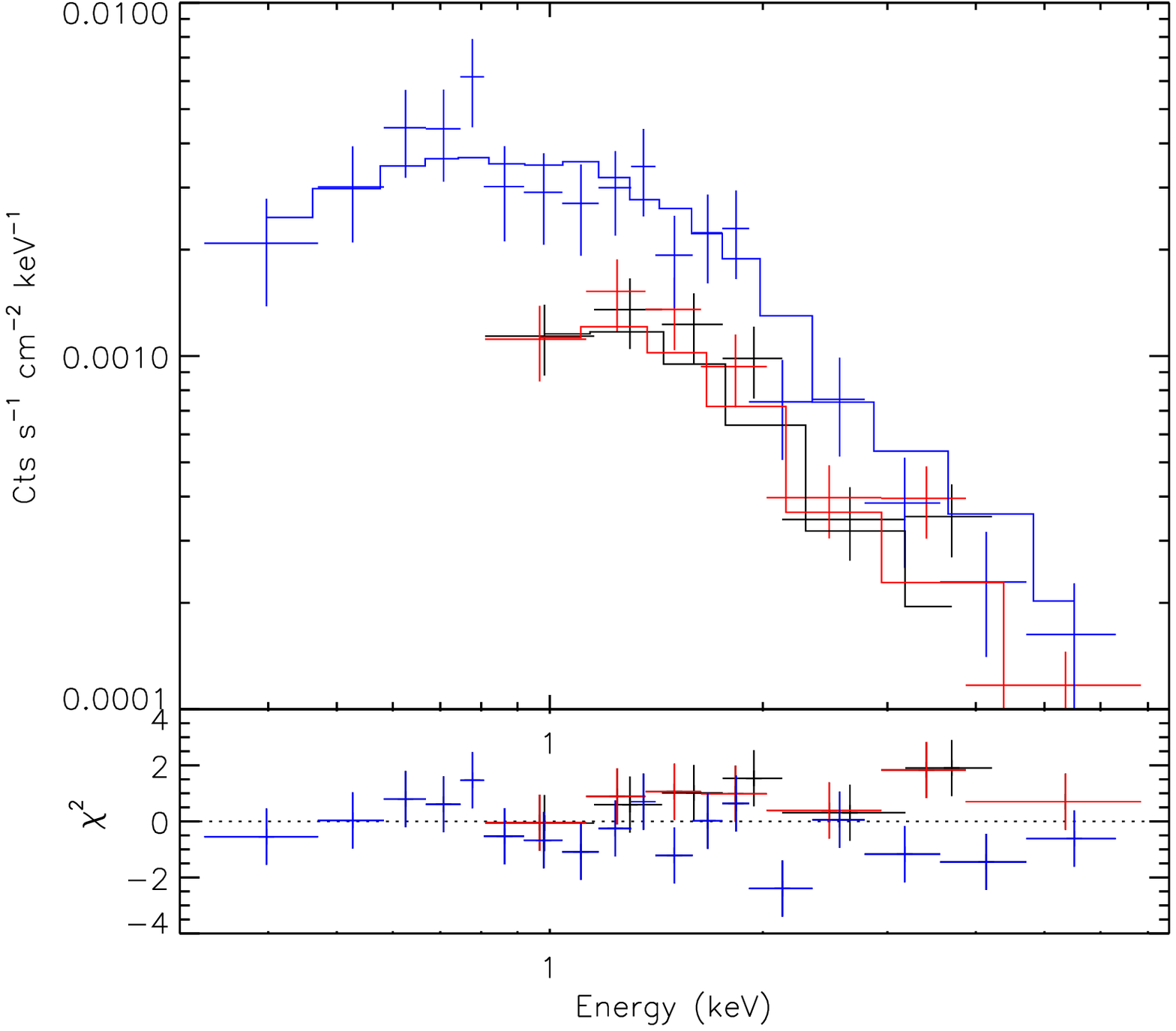}
\caption{\label{sx} PN and MOS spectra of the X-ray source at $\sim 12\arcsec$ from \hd 
obtained during the 2007 observation. The best fit model is an absorbed APEC thermal model. }
\end{center}
\end{figure}

{\ The X-ray images in 2007 and 2009 show a source $\sim 12\arcsec$ to the south of \hd A.
The source is brighter in 2007 than in 2009 (see Fig. \ref{isx}). 
The spectra obtained in 2007 are shown in Fig. \ref{sx}, those from 2009 observation
have poorer statistics and are not shown here. 
The source reveals a quite hard spectrum and, given that \hd A has a quite
soft spectrum, the contamination from this source to \hd A spectrum is negligible.
We have tried best fit with models composed by an absorbed power law and an absorbed 
APEC thermal model. In both cases the absorption is high:  $N_\mathrm H \sim 1.4\times 
10^{21}$ cm$^{-2}$.
Given that the spectra from 2009 have poor statistics, a simple model with an absorbed power 
law of index of $\sim1.7$ has a good agreement with the data. On the other hand, 
the spectra of 2007 have better statistics and the shape 
reveals some features around 1~keV so the simple power law is not appropriate.
When modeled with an APEC thermal component we find a high plasma temperature value, 
of order of 8 keV. The flux in 0.3--8.0 keV band is $1.9\times 10^{-13}$ erg s$^{-1}$ cm$^{-2}$ 
in 2007, whereas it is $1.4\times 10^{-13}$ erg s$^{-1}$ cm$^{-2}$ in 2009. 
The nearest optical / infrared counterpart to this X-ray source is an object in 
the 2MASS \citep{Cutri2003} and NOMAD \citep{NOMAD} catalogs at $\sim3\arcsec$. 
It has a poorly determined 2MASS photometry given the proximity to \hd\  and quite large 
proper motions that hint that the cataloged object should be quite close. 
On the other hand, the value of absorption of the X-ray spectrum of this source suggests that 
it should be quite distant from the Sun but still inside the Galaxy. In fact, typical values 
for the total N$_\mathrm H$ column in the direction of \hd\ are between $3.8\times 10^{21}$ 
cm$^{-2}$ and $3.9\times 10^{21}$  cm$^{-2}$ \citep{Kalberla05,Dickey90}.
By considering the flux in 2007 and 2009, and placing it at a distance of 1000~pc the 
luminosity yields $\log L_\mathrm X = 31.2 - 31.4$.
}

\section{\label{discussion}Discussion.}
{\  In this paper we have analyzed two X-ray observations of \hd A: the first 
obtained during a planetary transit on April 17 2007 and the second during a planetary 
eclipse occurred on May 18 2009. 
Here we discuss the features observed in the light curve of 2009 and we relate them to possible
effects of interaction of the planet and the host star.
\subsection{Observed features}
The first feature we have observed in 2009 is a {\em softening} of the emission during the 
planet eclipse  at a level of $\ga3\sigma$. 
Although similar variations of the spectrum hardness are seen in the light curve, 
this softening has a direct correspondence with the eclipse timing and, in addiction,
during the eclipse the star has the softest emission even recorded in both 2007 and 2009 
observations. The mean PI in EPIC-PN in the 0.3--1.5~keV band is $700\pm10$~eV before the 
eclipse and after the flare that occur post-eclipse, while it is $660\pm10$~eV at the mid eclipse. 

An analogy between \hd b and Jupiter here may be appropriate.
Jupiter shows enhanced soft emission in X-rays coming from the equatorial region 
and due to the interaction of solar X-rays with the upper atmosphere of the planet 
\citep{Branduardi07}. Furthermore, a harder X-ray emission is detected near the poles of Jupiter 
and may be due to charge exchange mechanisms caused by precipitating ions of carbon, 
sulfur and oxygen \citep{Hui09,Branduardi07}.
The flux in 0.2-7 keV coming from Jupiter aurorae is $\sim 7\times 10^{-14}$ erg s$^{-1}$ cm$^{-2}$
that corresponds to a luminosity of $\sim5 \times 10^{15}$ erg s$^{-1}$. 
{\ By using simple scaling, taking into account the star-planet distance in \hd , 
the higher luminosity of the star with respect to the Sun ($10 \times L_{X,\odot}$)
and the distance to \hd\ we estimate that the power emitted by aurorae would be 
$L_\mathrm{X,aur} \sim 1.4 \times 10^{21}$ erg s$^{-1}$, corresponding to a flux to Earth of 
$f_\mathrm{X,aur} \sim 3\times 10^{-20}$, which is too faint to be detected by XMM-Newton, 
unless there are additional sources of energy caused by the magnetic field topology. 
If the softening is related to the secondary eclipse it cannot be related to a scaled
emission of Jupiter for the case of \hd b.
With only one observation we are not in the position to assess 
that the softening is related to the secondary eclipse. A series of observations at the 
same phase would reveal if the softening is systematically recurrent.}

The second relevant feature present in the light curve of 2009 
is the most intense flare recorded in both observations, occurring 
$\sim$3~ks after the secondary eclipse. The increase of flux is equal to $\sim$85\% 
of the emission before the eclipse and the plasma temperature shows
a modest heating passing from $\sim$0.4 keV prior to the event to $\sim0.9$ keV. 
The properties of the flare like temperature, luminosity, decay rate
are similar to other flares observed in solar type stars without hot Jupiters.
During the flare the lines of OVII at $\sim22\AA\ $ are more luminous than before 
and after the event. 
The OVII complex remains in excess with respect to the best fit model of the flare  spectrum. 
High resolution RGS spectra  show that the $f$ OVII line remain 
quite luminous after the event and the $i$ line almost disappears. 

During its cooling phase, we see several impulsive events
(see Fig. \ref{smoothed}).  
It is as though there are multiple small flares occurring as the overall structure is cooling. 
The faster decay time would imply loops smaller than the main flaring loop.
A peculiar characteristic of the small impulses is that the rate increases while the
median PI steadily decreases.   
One way to understand this is to imagine that the flare occurred close to the limb, and
at least in occurrence of such events the corresponding part of the corona was emerging
into view.
If our interpretation of this evidence is correct, then the flare 
is occurring on the limb when the planet is at phase approximately 0.54. 
This would put the flare displaced by $\sim 77^o$ and the meeting the sub planetary point 
(see figure \ref{geometry}).

Given the timing of the flare one could speculate that the coronal region where it took 
place could be displaced by $\sim77^o$ and trailing the sub-planetary point.
The presence of spots and active regions on the stellar surface rotating synchronously with the 
planet have been reported in literature in stars with hot Jupiters. 
\citet{Shkolnik03} and \citet{Shkolnik05} report that HD~179949 and $\upsilon$~And show variability 
of chromospheric activity indicators phased with planet rotation hinting thus the presence 
hot spots rotating synchronously with the planet as a consequence of star-planet interaction. 
The spots are displaced with respect to the sub-planetary point by +70$^o$ and -170$^o$ 
respectively. Analogously to solar corona these spots should be the chromospheric footpoints 
of coronal loops.  
\citet{Shkolnik08} has suggested also some excess of flaring activity on \hd\ in phase with the planet 
rotation superimposed to the main long lasting active regions on the stellar surface.

One speculative hypothesis is that the large flare we have observed could arise from active regions
for which the activity is triggered by the complex magnetic interaction that should be present
between the star and the planet. 
In summary, we cannot draw a conclusion on this point with only one observation. 
As in the case of the softening, a series of further observations around the 
secondary eclipse could establish the systematic recurrence of high activity just after the end of 
the eclipse and coming from active regions triggered by star-planet interaction.

\subsection{Comparison with SPI models}
\citet{Lanza08} proposed an analytical model for the stellar magnetic field in the case of SPI. 
This model was aimed to explain the phase shifts 
of active regions with respect to the sub-planetary point observed.
With a suitable choice of the model parameters the author shows that this simple model 
can reproduce these phase shifts for \hd\, HD 179949, $\nu$~And and $\tau$ Bootis.
The prediction from this model for the case of  \hd\ is that a phase shift between sub-planetary point 
and active region should be $\sim78^o$. With the same model \citet{Lanza09} has addressed
the energy budget, the intermittent nature of the visibility of spots and the effect that
could be seen in radio and X-ray bands. From a qualitative point of view 
the stellar magnetic field should be able to drag material evaporating from the planet and 
to deposit it in a azimuthal flux rope in the outer part of the corona. 
This region should be the site of an enhanced X-ray activity
due to magnetic reconnection (see Fig. \ref{geometry} for a simple schematic). 
By comparing our observational results to this model, we observe that 
the large flare in 2009 can fit consistently  with this model and the
excess of flaring activity suggested by \citet{Shkolnik08} under the hypothesis
that it come from active regions displaced by $\sim75^o-78^0$ with respect to 
the sub-planetary point.
  
We have carried out preliminary MHD simulations of the star+planet
system within the dipolar field setup (``Case A'') of \citet{Cohen09},
but with the stellar and planetary parameters of the HD 189733 system.
The planetary magnetic field is taken to be half that of Jupiter
(with an equatorial field of 2 G); this is a reasonable assumption
since tidal effects will have reduced the planetary rotation compared
to that of Jupiter. 
Figure \ref{mhd} shows one frame of the time evolution of the magnetic field 
of \hd A+b system based on our MHD simulation. 
The simplified dipole model produces a density enhancement in the 
intervening interplanetary medium that is threaded by complex magnetic fields.
There is thus a greater potential to observe transients occurring in this
region.  The density enhancement is formed by plasma trapped into closed magnetic field lines 
that would have been opened in the absence of the planet interaction.  

In summary, we find elements of consistency between the observed features in X-ray band,
optical chromospheric indicators and model predictions. 
The magnetic interaction of planet and host star can enhance the activity in some regions 
of the stellar surface and that probably these are among
the  most active regions of the star chromosphere and corona. 
As suggested by \citet{Shkolnik08} basing on chromospheric activity indicators, the spot activity 
can be enhanced or reduced due to the complex magnetic field evolution. 
This part of the star-planet interaction is observed as flares. 
On the day side of the planet, it is plausible that the interaction manifests 
itself as colliding wind from the star on to the inflated atmosphere and with polar aurorae
that emit X-rays harder than the main coronal emission, although this emission remains 
too faint to be detected. 

Although our data do not allow us to draw specific conclusions, 
they suggest that effects of the magnetic
interaction between a star and a close planet could be detected in X-rays. 
A series of observations in the same phase range near the secondary eclipse and perhaps coordinated 
with optical observations of chromospheric activity indicators can allow to distinguish systematic effects 
caused by planet presence from random variability intrinsic to the coronal activity. 

\subsection{The age of \hd\ and its activity}
The age of \hd\ has been estimated by \citet{Melo06} to be greater than 0.6 Gyr basing on
indicators of chromospheric activity calibrated on normal Main Sequence stars. 
Furthermore these authors reported a very low abundance of Li (A(Li)$<-0.1$)
in \hd A. For comparison in the Sun  A(Li) is $\sim 2$ at an age of 4.6~Gyr. 
Lithium is destroyed during Main Sequence phase because of convection 
that brings lithium in the inner layers of the star where at a temperature of 2--3 MK is burned. 
Both Li and chromospheric indicators are sensitive for ages $< 0.5-1$ Gyr. 
In the case of \hd\ the use of chromospheric lines could be biased by the enhanced activity 
of the star itself and, eventually, by SPI. 
Our X-ray observation suggests a better constraint on its age.

The non-detection of \hd B, a M4 spectral type star, at a level of $L_\mathrm X \le 10^{27}$ 
erg s$^{-1}$ suggests an age estimate of \hd\ older than the current one. 
M-type stars exhibit a decay of X-ray emission during Main Sequence phase slower than in G and K type
stars due to the lower rate of dissipation of angular momentum and the link between rotation and 
dynamo efficiency.
\citet{Feigelson04} have studied a sample of low mass stars in the {\em Chandra Deep Field North}.
In this sample M4 stars are characterized by $L_\mathrm X \sim 10^{28}$ erg s$^{-1}$, and the 
age of the sample is presumed to be greater than 3~Gyr. 
Given the non-detection of \hd B we suggest that the age of this system could 
be tentatively estimated older than 2 Gyr.

Placing the system at an age of 2~Gyr the X-ray luminosity of \hd\  ($L_\mathrm {X,*} \sim 10 
L_\mathrm{X, Sun}$) fits the trend observed from studies of X-ray luminosities and activity in 
K-type stars of open clusters and field stars \citep[see][and references therein]{Giardino08}. 
\citet{Mamajek08} consider the age estimates from various indicators like Ca H\&K
lines, A(Li), rotation and X-ray emission, obtaining thus several relations between age and
these observables. With a rotational period of $\sim12$d and a K1 type star, \hd\ would fit close
to the age of Hyades \citep[see their fig. 10]{Mamajek08}. But the value of 
$\log R^\prime_{HK}$ reported by \citet{Melo06} and the curves given by \citet{Mamajek08}
in their fig. 12 would suggest an age of $\tau \ge 1.7$ Gyr. 
Furthermore, U, V, W space motion components (U = -8.6 km/s, V = -14.4 km/s, W = 015.9 km/s) 
are not compatible with those of Hyades moving group \citep{Montes2001}.
In conclusion, although the determination of the age of an isolated star can be difficult and poorly 
constrained, the non-detection of the M type companion below $\log L_\mathrm X \sim 27$ hints an age 
of \hd\ around 1.7--2 Gyr.}

\section{\label{conclusions} Conclusions}
{ \
In this paper, we have presented the analysis and the interpretation of X-rays light curves and
spectra that we have acquired during a secondary eclipse of the planet \hd b observed in 2009
with the XMM-Newton satellite. For completeness we have also analyzed in a homogeneous way a previous
observation taken in 2007 during a planetary transit on the same target. Our group has carried also
MHD simulations of the magnetic interaction between star and planet. 
From the observations we have estimated that the average X-ray flux of the star in 2009
($f_\mathrm X \sim 3.7\times 10^{-13}$ erg s$^{-1}$ cm$^{-2}$) has been has been 45\% higher 
than the average flux in 2007.
The light curve of 2007 shows impulsive variability which is considered quite usual for a K-type star
quoted at an age of 1~Gyr. During the planetary transit we do not observe spectral or rate variations
timed with the transit. 

In the light curve of 2009 observation we have observed a softening  of the spectrum 
significant at a $3\sigma$ level during the secondary eclipse and the biggest flare recorded in 
both observations occurring 3~ks after the end of the planet eclipse. The softening is well phased 
with the planet eclipse. The median of the spectrum change from an average value
of $700\pm10$ eV before and after the eclipse to $650\pm10$ eV at the eclipse center.
The decay phase of the main flare in the light curve shows several superimposed minor 
ignitions perhaps occurring in the same region where the main flare took place. 
We have fitted the spectrum subtracted from the pre-flare spectrum with a 
mono-thermal APEC model. The flare exhibits a average plasma temperature of 0.9 keV 
while the quiescent plasma temperature is around 0.5 keV.
The RGS spectrum shows a variability of the relative strength of the lines of OVII triplet
prior, during and after this flare.
In particular we noticed the reduction of inter-combination and forbidden lines during the flare,
and a bright forbidden line after the end of the flare. These two behaviors of the OVII triplet 
lines are explained as an increase of density in the flaring region during the ignition, and 
a persistent low density bright/hot plasma after the end of the flare. 
By using a simple model of cooling loop as in \citet{Serio1991} and \citet{Reale1997}, 
we infer a size of the flaring region of order of the stellar radius. 
From $L_\mathrm X \sim v_\mathrm{rot}^2$ scaling law 
\citep{Pallavicini81} we infer a magnetic field twice the strength of the solar one.

In order to explore the possibility that the planetary magnetosphere can directly affect 
the stellar coronal structure, we have conducted MHD simulations of a model of interacting magnetic 
fields with opposite alignment as in Case A of \citet{Cohen09}.
From these simulations we infer that due to
the relative motion of the planet and the star the global magnetic field is distorted and
has an enhanced probability of reconnection events that can occur on the space between 
planet and star. Given the coupling of magnetic field and plasma, an enhanced plasma density 
is present around the planet and it forms a trail rotating synchronously with the planet. 
Globally we find consistency between observations and predictions from MHD simulations, hinting
that the overall activity could be enhanced by magnetic SPI. Hence, the age estimates of the system
using activity indicators are suspect. Currently the age of \hd \ is estimated to be at 
least $\ge 0.6$ Gyr. The M-type companion, \hd A, is not detected at a level of luminosity of 
$L_\mathrm X < 10^{27}$  erg s$^{-1}$ suggesting thus an age for the system older than 1.7--2~Gyr. 

We have also analyzed the spectrum of a X-ray source close to \hd\ and brighter in 2007 than in 2009.
The spectrum is quite hard and from best fit we derive an absorption of 
$N_\mathrm H \sim 1.4\times 10^{21}$ cm$^{-2}$ suggesting that this is a object in the Galaxy but quite distant
from the Sun. 
}

\begin{acknowledgements}
The XMM guest investigator program supported IP through grant NNX09AP46G. 
SJW was supported by NASA contract NAS8-03060 to the Chandra Science Center. 
VLK acknowledges CXC NASA contract NAS8-39073, OC acknowledges NASA LWSTRT Grant NNG05GM44G. 
\end{acknowledgements}

% \bibliographystyle{aa}
% \bibliography{bib_latex}

\begin{figure}
\begin{center}
\includegraphics[width=0.49\columnwidth,angle=0]{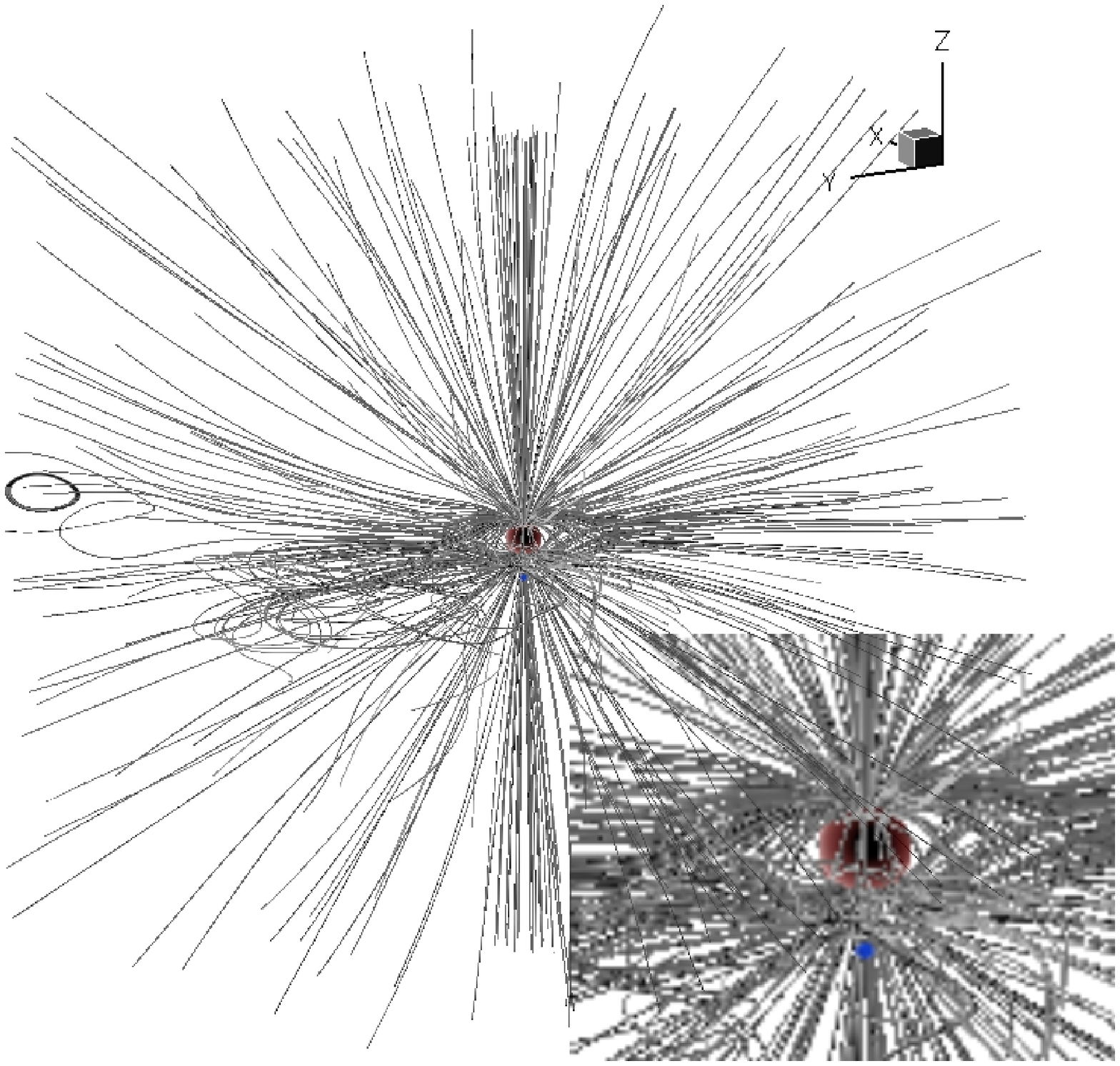}
\includegraphics[width=0.49\columnwidth,angle=0]{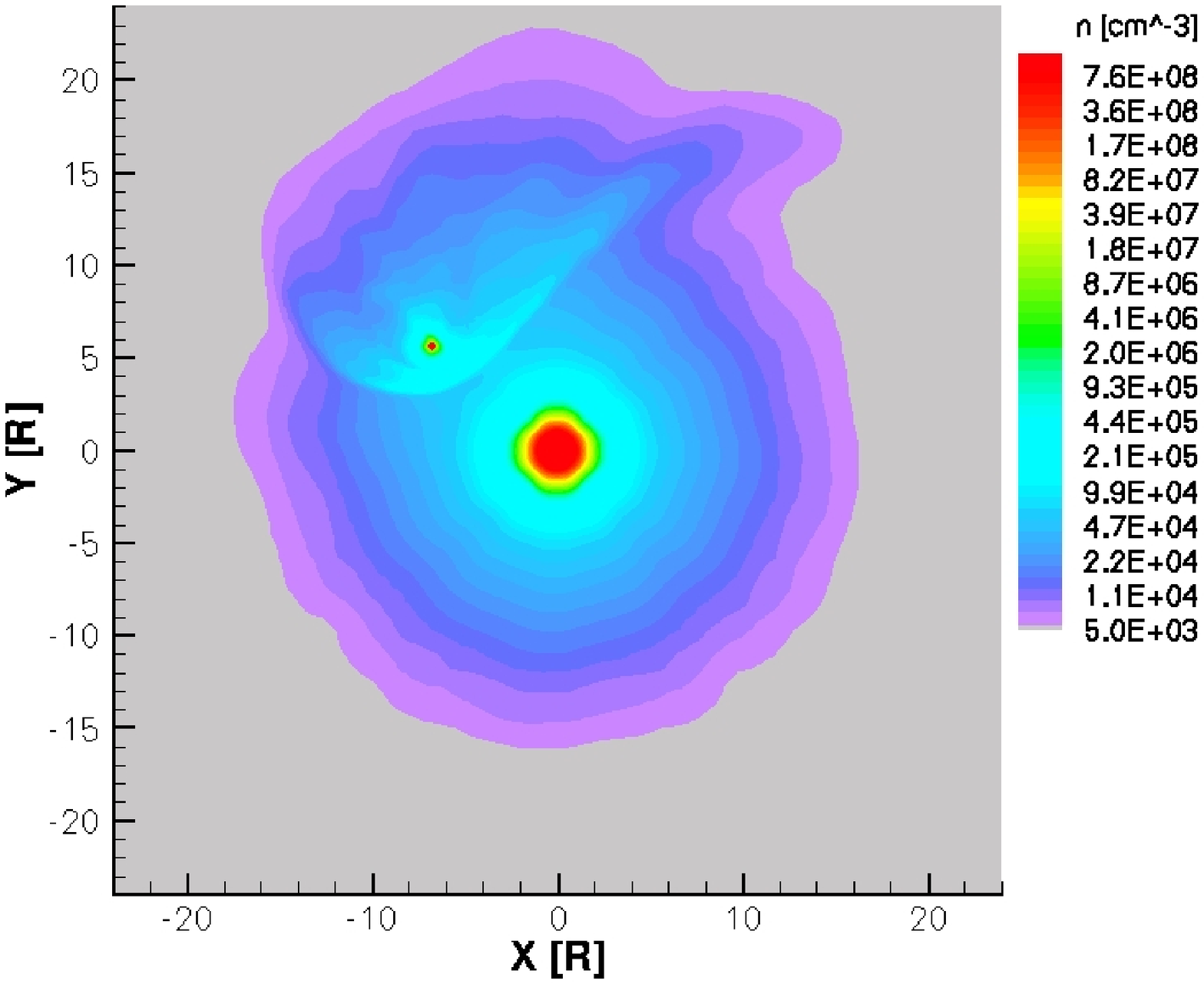}
\end{center}
\caption{\label{mhd} {\ MHD model of the star+planet. On the left panel: 
one frame of the MHD simulation. 
The star and the planet are represented by the red and blue spherical shades, respectively.
Lines of the magnetic field are drawn in gray, the inset plot in the right bottom part show
the central star and the planet. On the right panel: density image viewed from the top. 
Model predict a denser trail orbiting synchronously with the planet. 
}
}

\end{figure}

\begin{table}
\caption{\label{spec} Best fit parameters of spectra collected during the observations 
in 2007 and 2009. For 2009 we list the results of the three time intervals: before flare, 
during flare and after the flare.}
\small
\begin{tabular}{l l l l l l l l l} \tableline \tableline \\
Dates &  kT~1 & E.M. 1 & kT 2 &  E.M. 2  & $\chi^2_{red}$ & P($\chi^2 > \chi^2_0$) & Flux  & Notes \\ \tableline
      & keV     & 10$^{51}$ cm$^{-3}$ & keV & 10$^{51}$ cm$^{-3}$  &  &  & 10$^{-13}$ erg s$^{-1}$ cm$^{-2}$ & \\
2007  &  0.57$_{0.51}^{0.59}$ &  0.95$_{0.76}^{1.26}$ & 0.24$_{0.16}^{0.25} $ &0.90$_{0.73}^{1.13}$ & 1.13&  0.14 &    2.50   & whole obs. \\   
2007  &  0.59$_{0.48}^{0.65}$ &  2.50$_{1.98}^{3.00}$ & -- & --&  1.45 & 0.07  &   3.14   & first flare \\
2007  &  0.56$_{0.48}^{0.59}$ &  0.95$_{0.76}^{1.40}$ & 0.24$_{0.18}^{0.27} $ &0.97$_{0.64}^{1.21}$ & 1.16&  0.08 &    2.50   & no first flare \\   
2009  &  0.42$_{0.40}^{0.46}$ &  2.81$_{2.46}^{3.15}$ & -- & -- & 1.29 & 0.04  &   3.11   & Pre flare \\   
2009  &  0.61$_{0.57}^{0.63}$ &  3.15$_{2.79}^{3.51}$ & -- & -- & 1.37 & 0.02  &   4.38   & flare    \\   
2009  &  0.50$_{0.46}^{0.53}$ &  3.22$_{2.83}^{3.63}$ & -- & -- & 0.87 & 0.78  &   3.72   & Post flare \\   
2009  &  0.86$_{0.77}^{1.02}$ &  0.32$_{0}^{0.64}$    & -- & -- & 1.1  & 0.25  &   1.19   & Flare -- Pre Flare  \\   
\tableline
\end{tabular}
\end{table}

\end{document}